\documentclass[twocolumn,aps,pra]{revtex4}
\usepackage{amsmath,amssymb}
\usepackage{graphicx}
\usepackage{epstopdf}
\usepackage[normalem]{ulem}
\usepackage[usenames]{color}

\newcommand{\be}{\begin{equation}}
\newcommand{\ee}{\end{equation}}
\newcommand{\bk}{{{\bf{k}}}}

\newcommand{\bQ}{{{\bf{Q}}}}
\newcommand{\br}{{{\bf{r}}}}

\newcommand{\bq}{{\bf{q}}}
\newcommand{\hx}{{\hat{x}}}
\newcommand{\hy}{{\hat{y}}}

\newcommand{\bea}{\begin{eqnarray}}
\newcommand{\eea}{\end{eqnarray}}
\newcommand{\ra}{\rangle}
\newcommand{\la}{\langle}

\newcommand{\upa}{\uparrow}
\newcommand{\dna}{\downarrow}

\newcommand{\dg}{{\dagger}}
\newcommand{\pdg}{{\phantom\dagger}}

\renewcommand{\vec}[1]{\mathbf{#1}}

\begin{document}

\title{Anisotropic quantum quench in the presence of
frustration or background gauge fields: A probe of bulk currents and topological chiral edge modes}

\author{Matthew Killi$^1$, Stefan Trotzky$^{1,2}$, and Arun Paramekanti$^{1,3,4}$}
\affiliation{$^1$Department of Physics, University of Toronto, Toronto, Ontario, Canada M5S 1A7}
\affiliation{$^2$Center for Quantum Information and Quantum Control, University of Toronto, Toronto, Ontario, Canada M5S 1A7}
\affiliation{$^3$Canadian Institute for Advanced Research, Toronto, Ontario, M5G 1Z8, Canada}
\affiliation{$^4$Kavli Institute for Theoretical Physics, University of California, Santa Barbara, CA 93106, USA}
\begin{abstract}
Bosons and fermions,
in the presence of frustration or background gauge fields, can form
many-body ground states that support equilibrium charge  or spin currents.
Motivated by the experimental creation of frustration or synthetic gauge fields in
ultracold atomic systems,
we propose a general scheme by which making
a sudden {\it anisotropic} quench
of the atom tunneling across the lattice and
tracking the ensuing density modulations
provides
a powerful and gauge invariant route to probing diverse equilibrium current patterns.
Using illustrative examples of trapped superfluid Bose and normal Fermi systems in the
presence of artificial magnetic fluxes on square lattices, and frustrated bosons in a triangular lattice, 
we show that this scheme to probe equilibrium {\it bulk} current order
works independent of particle statistics. We also show that such quenches can detect chiral
{\it edge} modes in gapped topological states, such as quantum Hall or quantum spin Hall insulators.
\end{abstract}

\date{\today}

\maketitle
The physics of fermions or bosons moving in background gauge fields is of great interest in various condensed matter systems such as quantum Hall liquids \cite{stone.book},
topological insulators \cite{TI.review}, quantum spin liquids \cite{balents}, and the cuprate
superconductors \cite{leewen.rmp}.
Such abelian or non-abelian gauge fields, imposed externally or generated by strong correlation
effects, can result in equilibrium
charge or spin currents of electrons.
For instance,
a type-II superconductor in
a magnetic field forms an Abrikosov vortex lattice that supports a periodic 
bulk current pattern formed by Cooper pairs swirling around each vortex \cite{abrikosov}.
A uniform magnetic field for lattice electrons
can lead to
topologically nontrivial states with a quantized
Hall conductance and chiral edge currents \cite{tknn}. Such electronic charge currents in a solid 
produce their own characteristic magnetic fields, and
can thus be probed by using magnetic
microscopy or neutron scattering. These
tools have been used to study vortices in type-II superconductors \cite{magmicroscopy}, 
to search for complex current patterns in the high temperature cuprate 
superconductors \cite{greven}, or to look for edge currents in purported chiral superconductors 
such as SrRu$_2$O$_4$ \cite{chiralsc}.
Electronic spin currents in solids, by contrast, are
harder to measure.
A direct observation of the spin Hall effect in semiconductors 
involves driving a charge transport current and optically detecting the
spin accumulation at the transverse edges of the sample \cite{awschalom.2004}.
Observing equilibrium spin currents is a more difficult challenge; only recently
have experiments shown that the quantum spin Hall edge modes in
two-dimensional HgTe quantum wells carry spin polarization \cite{molenkamp2012}.

Over the past few years, experiments in the field of ultracold atomic gases have also begun to study the
effects of ``artificial'' orbital magnetic fields \cite{spielman.nature2009,bloch.prl2011,dalibard.rmp}
and spin-orbit coupling \cite{yjlin.nature2011} in the hope of creating new states of atomic
matter. These experiments can potentially realize various topological phase transitions and a wide variety of 
states with
equilibrium mass currents \cite{MD.prl2009,MD.njp2010}. Such mass currents also arise in the presence of `lattice shaking'
\cite{sengstock.science2011,sengstock.prl2012}, the combination of Raman lasers and radio frequency fields \cite{spielman.prl2012}, or from populating higher optical lattice bands with
bosons \cite{hemmerich.prl2011,hemmerich.nphys2011}, both of which lead to kinetic frustration and
possible spontaneous time-reversal symmetry broken superfluids \cite{tosi.prl2005,dassarma.prl2008}.
Two-component bosons, in the presence of spin-orbit coupling and strong correlations, have recently
been proposed to support
Mott insulator states with complex magnetic textures, such as vortex crystals and skyrmion lattices
\cite{wscole.prl2012,radic.prl2012,congjunwu.pra2012}. Upon decreasing the Hubbard repulsion,
such Mott insulators transition into superfluids, which retain the magnetic textures, with the magnetic
order imprinting nontrivial Berry phases on the bosons and leading to intricate superfluid current patterns 
\cite{wscole.prl2012}. Spinless fermions with longer range repulsive interactions and frustrated hopping
on the triangular lattice have also been recently proposed to realize states with spontaneously broken 
time-reversal symmetry and loop currents \cite{eckardt.2012}.

But, \emph{how can we experimentally deduce the equilibrium mass current patterns for such neutral
atomic gases}? This is rapidly becoming an important issue since cold atomic gases are poised to create
a number of interesting condensed matter states using such gauge fields. Experiments on bosonic atoms
use peaks in the boson momentum distribution to infer the location of the boson dispersion minima
induced by the presence of synthetic magnetic fluxes \cite{bloch.prl2011}; it would thus be extremely
valuable to have a complementary technique that directly probes gauge invariant equilibrium mass
currents induced by such background synthetic gauge fields or frustration.

In this paper, we argue that the study of density dynamics triggered by specific quantum quenches
provides a powerful route to probing atom mass currents, and we present an extended discussion of this
idea going well beyond our previous work \cite{short}. Our proposal to measure equilibrium atom currents
induced by the presence of a gauge field relies on measurements of the atom density, and is inspired by
the significant experimental progress in measuring even lattice scale density modulations. Many such
density mapping tools have been experimentally demonstrated in recent years, such as noise correlations
\cite{noisetheory,noiseexpt}, Bragg scattering \cite{kuhr.prl2011}, which is analogous to X-ray
scattering used to deduce the crystal structure of solids, and {\it in situ} microscopy
\cite{greiner.nature2009}, which is similar to scanning tunneling microscopy at a crystal surface in the sense 
that both probe real-space lattice scale physics.

\begin{figure}[t] 
	\includegraphics[width=0.48 \textwidth]{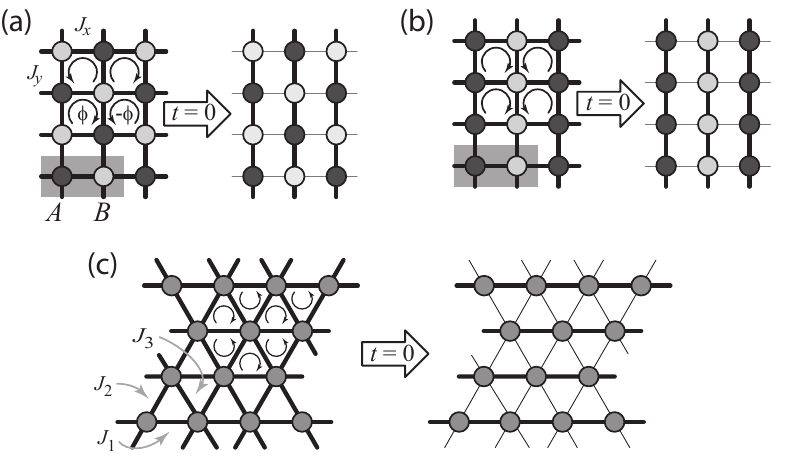} \caption{ Illustration of the investigated scenarios: (a) Unidirectional quenches of the tunnel coupling in a square lattice with staggered (checkerboard) magnetic flux, (b) unidirectional quenches in a square lattice with a striped flux pattern and (c) bidirectional quenches in a triangular lattice with frustrated hopping. Strong and weak tunnel couplings are illustrated by thick and thin lines and the  magnetic unit cell is highlighted. The quench dynamics for all three cases was investigated considering a Bose superfluid; case (a) was additionally studied for noninteracting fermions.}
\label{Fig:quench_schematics}
\end{figure}

One key idea we use is to make a specific quantum quench of the Hamiltonian that violates the steady state {\it divergence-free} condition on equilibrium currents \cite{short}. This causes an imbalance between currents entering and leaving different sites of the lattice. As dictated by the continuity equation, this leads to characteristic density build-up or depletion with a specific pattern across the lattice, which reflects the initial currents in equilibrium. A ``quasi-local'' current probe of this type has been used in a recent experimental study of nonequilibrium dynamics in a (one dimensional) 1D Bose gas \cite{trotzky}. In other cases, one can design suitable quenches that lead to spontaneous macroscopic dipolar density oscillations, corresponding to center of mass oscillations of the atom cloud in the harmonic trap. In either case, imaging the subsequent density variation across the lattice yields {\it real space} information about the initial currents. Thus, just as the usual time of flight images probe momentum information by studying real space atom positions after a time delay following release from a trap, our proposed scheme yields atom current information by converting them into density images after a time delay. Such quenches along with the underlying current patterns are schematically depicted in Fig.~\ref{Fig:quench_schematics} for some of the examples explored in this paper.
                                 	
Our proposed scheme has the additional advantage of being independent of particle statistics, and is applicable to both bosons and fermions, as we illustrate here for both Bose superfluids and degenerate Fermi gases. Moreover, it can also be used to probe both bulk currents and edge currents in the system. In addition to the systems explored here, we also expect that such quenches could also probe the current pattern in the recently studied chiral Bose Mott insulator \cite{dhar.pra2012,dhar.2012b}, 
and, more generally, the dynamics can also be used to study spin currents of atomic matter, since one can experimentally probe the spin-resolved density in the lattice as demonstrated in recent experiments \cite{weitenberg.nature2011}.

Finally, quantum quenches have long been of great interest in the context of such diverse and important issues as the approach to equilibrium in closed quantum systems, defect production induced by tuning the Hamiltonian across various quantum phase transitions, and extensions of scaling and renormalization group ideas to dynamics across quantum phase transitions such as in the context of the Kibble-Zurek problem \cite{nondyn, KZ.1,KZ.2,KZ.3,KZ.4,KZ.5,KZ.6,KZ.7}. Our work, thus, additionally serves to bring together these two threads of research --- synthetic gauge fields and quantum quenches --- by suggesting that studying quantum quenches and quench-induced dynamics in the presence of gauge fields would be a useful direction to pursue given the recent experimental and theoretical advances in these areas. Indeed, there have even been theoretical proposals to produce {\it dynamical} gauge fields in ultracold atomic systems \cite{buchler,lewenstein.2011}, and a recent suggestion that one could use quenches in Bose-Fermi mixtures to simulate `string breaking' dynamics in a model of fermions coupled to such dynamical gauge fields \cite{zoller.2012}.

A brief version of some of the results in this paper is contained in Ref.\ \cite{short}. Here, we outline a general scheme to extract currents, expand on the various analytical results for noninteracting cases and for interacting Bose systems, give further details on quenches for quantum Hall states, study a triangular lattice frustrated superfluid, and explore the effect of random phase fluctuations imprinted on the initial state prior to the quench in order to show that weak thermal fluctuations in a Bose superfluid do not affect our central conclusions.

\section{General scheme and models}

Consider a general $d$-dimensional lattice system that carries local currents $\vec{j}(\br)$ in equilibrium. In order to uncover a specific component of the current, say $j_x = \vec j \cdot \vec{\hat{x}}$, let us make a quench of the Hamiltonian at time $t=0$ that instantaneously turns off the hopping along all transverse directions. The currents $\vec j(\vec r)$ remain unchanged at the
instant of the quench; the quench itself induces no extra currents. However, at all subsequent times, the density and current will evolve in an effectively one-dimensional system along the $x$ direction. The continuity equation following the quench takes the form
\bea
- \frac{\partial n(\br,t)}{\partial t} = j_x(\br+\hat{\vec x}/2) - j_x(\br - \hat{\vec x}/2),
\eea
where the right-hand side denotes the lattice divergence of the current.

\subsection{Short time analysis}
If we focus on short times after the quench, the change in density $\delta n(\vec r, t)$ will be dictated by the initial equilibrium currents in the unquenched direction, independent of the details of the Hamiltonian or the statistics of the particles making up the fluid. Therefore, at short time $\epsilon$ after the quench and in Fourier space, we can set $\Delta n(\bq,\epsilon) \approx  - 2 i \epsilon \sin (q_x/2)  j_x(\bq,0)$, with $j_x(\vec{q},0)$ being the Fourier transform of the pre-quench equilibrium current. For this
short time analysis to be valid, we must
choose $1/\epsilon$ to be comparable to the tunneling rate, but much shorter than the frequency of the subsequent density
oscillations discussed below.

For $q_x \neq 0$, we can invert this relation to obtain
\bea
j_x(\bq,0) \approx  - \frac{1}{2 i \epsilon \sin(q_x/2)}   \delta n(\bq,\epsilon).
\eea
Thus, a measurement of the excess density $\delta n$, which builds up shortly after the quench and decomposing it into its spatial Fourier components, corresponds to a determination of all the nonzero Fourier components of $j_x$.

To determine the Fourier components of the density, one could resort to tools such as Bragg scattering \cite{kuhr.prl2011}, noise correlation measurements \cite{noisetheory,noiseexpt,inguscio.2008} or superlattice aided band-mapping techniques \cite{sebbystrabley.2006}. Assuming simple current patterns, only a few Fourier components will be nonzero. For low filling $<1$, in situ imaging after freezing out all atom hopping provides the most direct measurement \cite{greiner.nature2009, sherson:2009}. For larger filling, however, additional effort has to be made to overcome the parity mapping inherent to this method.

For $q_x = 0$, the above inversion fails. This component, however, corresponds to a uniform current offset and can be detected in the presence of an external trapping potential by monitoring the change in the center-of-mass position after short times $\epsilon$.

Aside from the short-time dynamics following a complete quench, it is also useful to study the long-time dynamics and the dynamics following a weak quench; both these issues are amenable to analysis and experiments, and are interesting in their own right. At the very least, such an analysis of the dynamics allows us to address the issue of how long the system needs to evolve before the measurement to ensure there is an experimentally measurable density accumulation. In order to make progress on this front, we focus on specific model Hamiltonians.

\subsection{Models}
We are interested in applying the general scheme outlined above to uncover the underlying current patterns of interesting and experimentally relevant examples of many body states of bosons and fermions. We therefore focus on the following models of Bose superfluids: (\textit{i}) bosons on a square lattice in a staggered, checkerboard-like magnetic flux pattern as shown in Fig.~\ref{Fig:quench_schematics}(a), (\textit{ii}) bosons on a square lattice with a striped magnetic flux pattern as realized in Ref.~\onlinecite{bloch.prl2011} (Fig.~\ref{Fig:quench_schematics}(b)), and (\textit{iii}) a triangular lattice model of frustrated bosons moving in a staggered flux
pattern as realized in Ref.~\cite{sengstock.science2011} (Fig.~\ref{Fig:quench_schematics}(c)). In addition, we study two models of noninteracting fermions: (\textit{iv})
non-interacting fermions on a square lattice in a staggered magnetic flux (Fig.~\ref{Fig:quench_schematics}(a)), and (\textit{v}) the Hofstadter model of fermions with a uniform flux on a square lattice, which results in an integer quantum Hall phase. 
As depicted in Fig.~\ref{Fig:quench_schematics}, we study sudden quenches where we turn off (or weaken) the hopping along all directions except one. For the cases (\textit{i}), (\textit{ii}) and (\textit{iv}), the quench we study corresponds to turning off or weakening the hopping along one direction, leading to sublattice-density oscillations. In case (\textit{iii}), the quench turns off the hopping along two of the three bond directions, resulting in macroscopic dipole oscillations of the atom cloud in a trap. 
For the quantum Hall case (\textit{v}), a unidirectional quench is shown to lead to quadrupole oscillations dominated by chiral edge currents.

In order to model the density dynamics of these systems, we resort to two approaches. For bosons, we study examples with repulsive contact interactions modelled by a Hubbard Hamiltonian of the schematic form
\be
H_{\rm BH} \! =\! - \sum_{\br,\br'} J^\pdg_{\br,\br'} b^\dg_\br 
b^\pdg_{\br'} + \sum_\br V^\pdg_\br b^\dg_\br b^\pdg_\br + \frac{U}{2} \sum_\br b^\dg_\br b^\dg_\br b^\pdg_\br
b^\pdg_\br.
\label{eq:BHM}
\ee
Here, the complex hopping amplitudes (transfer integrals) $J^\pdg_{\br,\br'}$ encode the artificial fluxes, $U$ is the on-site Hubbard repulsion, and $V^\pdg_\br = V_0 (x^2+y^2)$ is a harmonic trap potential. We analyze these boson models using a Gross-Pitaevskii (GP) approach, which replaces $b^\pdg_\br$ by a ``condensate wavefunction'' $\Psi^\pdg_\br$. This leads to an equilibrium energy functional
\begin{eqnarray} 
E_{\rm GP} = - \sum_{\br, \br'}
J_{\br,\br'} \Psi^*_{\br} \Psi^\pdg_{\br'} \!+\! \sum_\br V_\br |\Psi_\br|^2 \!+\! \frac{U}{2} \sum_\br |\Psi^\pdg_\br|^4 \,,
\label{GPfunc}
\end{eqnarray}
which must be minimized to obtain the initial equilibrium superfluid ground state. Starting from this ground state of the pre-quench Hamiltonian, we suddenly decrease the
tunneling amplitudes from their initial values, $(J_x^i , J_y^i)$, to their final values, $(J_x^f , J_y^f)$, at time $t = 0$ and study the subsequent time evolution of this state. The quench-induced dynamics is then obtained by solving the time dependent GP equation
 \be 
 i \hbar \frac{\partial \Psi_\br(t)}{\partial t} =
- \sum_{\br'} J_{\br,\br'}^f \Psi_{\br'}(t) + [U |\Psi_{\br}(t)|^2 \! +\! V_\br] \Psi_\br(t). 
\label{Time}
\ee
Henceforth, we set $\hbar=1$. While we present an analytical discussion of the quench-induced dynamics in uniform systems, we also present numerical solutions for Bose superfluids in a harmonic trap. Specifically, the equilibrium state is obtained by 
numerically minimizing the GP energy functional, while the post-quench dynamics is obtained by numerically solving the time-dependent GP equation. Details of the numerical procedures are contained in Appendices A and B.

For fermions, we restrict ourselves to noninteracting (spinless) examples for which we can diagonalize the Hamiltonian either analytically or numerically for large systems. These fermion Hamiltonians schematically take the form
\be
H = - \sum_{\br,\br'} J^\pdg_{\br,\br'} 
f^\dg_\br f^\pdg_{\br'} + \sum_\br V^\pdg_\br f^\dg_\br f^\pdg_\br.
\ee
We again imagine quenching the fermion hopping at time $t=0$, with the time evolution being governed by the time-dependent Schr\"odinger equation. Knowing all the pre-quench and post-quench eigenstates and eigenvalues is then sufficient to reconstruct the dynamics of various observables after the quench.

\section{ Bose superfluid in a staggered magnetic flux background}

We begin by studying a weakly interacting superfluid of bosons on a 2D square lattice,
described by the Bose-Hubbard model Eq.~(\ref{eq:BHM}). 
We take $J_{\br,\br'} \! = \! J^*_{\br',\br} \! \neq \! 0$ only for nearest neighbors, and choose $J_{\br,\br+\hat{\vec x}} \! = \! J_x$ real 
and $J_{\br,\br + \hat{\vec y}} \! = \! J_y \exp [i (-1)^{x+y}\phi/2]$. This yields staggered magnetic fluxes, $\pm \phi$, that pierce the elementary square plaquettes in a checkerboard pattern [see Fig.~\ref{Fig:quench_schematics}(a)]; a route to realizing such a flux pattern has been proposed previously \cite{hemmerich.stagsf}.

\subsection{Equilibrium state in the absence of a trap}

For weak interaction, $U \! \lesssim \! J_x,J_y$, we solve for the equilibrium ground state by minimizing the GP energy functional Eq.~(\ref{GPfunc}) \cite{GPE}. In the absence of a trap ($V_\br=0$) we first diagonalize the kinetic energy in the Hamiltonian~(\ref{eq:BHM}). In momentum
space, the kinetic energy takes the form
\bea
H = \sum_{\bk \in RBZ}
\begin{pmatrix}  \Psi^\dg_\bk& \Psi^\dg_{\bk+\bQ} \end{pmatrix}
(\varepsilon_\bk \tau^z + \gamma_\bk \tau^y)
\begin{pmatrix}  \Psi^\pdg_\bk \\ \Psi^\pdg_{\bk+\bQ} \end{pmatrix}
\eea
where $\tau^{y,z}$ are Pauli matrices, $\bQ=(\pi,\pi)$, and $RBZ$ denotes the reduced Brillouin zone due to the unit cell doubling resulting from the flux. Here, we have defined
\bea
\varepsilon_\bk &=& -2 J_x \cos k_x - 2 J_y \cos(\phi/2) \cos k_y \\
\gamma_\bk &=& - 2 J_y \sin(\phi/2) \cos k_y\,.
\eea
Let us restrict ourselves to flux values $0 < \phi < \pi$. We find that the minimum eigenvalue, $ \lambda_\bk = - \sqrt{\varepsilon_\bk^2  + \gamma^2_\bk}$, then occurs at $\bk=(0,0)$. For $J_x=J_y=J$ in equilibrium, this is given by 
\be
\lambda_0 = - 4 J \cos \frac{\phi}{4}\,.
\ee
The GP wavefunction in the absence of a trap is given by the wavefunction corresponding to this minimum eigenvalue, $\Psi_\br = \sqrt{n_0} (u_0 + i v_0\eta_{\vec r})$, where
\bea
u_0 &=& \frac{1}{\sqrt{2}} \left(1+ \frac{\varepsilon_0}{\lambda_0}\right)^{1/2}\,,
\label{equ0} \\
v_0 &=& \frac{1}{\sqrt{2}} \left(1- \frac{\varepsilon_0}{\lambda_0}\right)^{1/2}\,,
\label{eqv0}
\eea
and $\eta_{\vec r} \equiv (-1)^{x+y}$. In this initial equilibrium state, the density $|\Psi_{\vec{r}}|^2 = n_0$
is uniform and there is an alternating checkerboard pattern of circulating currents on the elementary square plaquettes. The magnitude of this staggered current is given by $4 J u_0 v_0 n_0$ on each bond.

Since the density in this state is uniform, this wavefunction continues to be the ground state of the full GP equation in the absence of a trap. In later subsections where we present a numerical solution to the GP equation in the presence of a trap, the currents and densities are nonuniform.

\subsection{Exact analysis of a quench for noninteracting bosons with no trapping potential}

For noninteracting bosons, it is simple to analyze the quench dynamics in the absence of a trap, since we explicitly know the energies and eigenstates before and after the quench. Specifically, let the equilibrium time-independent wavefunctions in the pre-quench and post-quench Hamiltonians be given by $\sqrt{n_0} (u_0 + i v_0 \eta_\br)$ and  $\sqrt{n_0} (\tilde u_0 + i \tilde v_0 \eta_\br)$ respectively, where the coefficients of the uniform and staggered components
are determined from Eq.~(\ref{equ0}) and Eq.~(\ref{eqv0}). 
We can then write the post-quench time-dependent wavefunction in the form 
\bea
\Psi(\br,t) &=& \sqrt{n_0} \left[ \alpha (\tilde u_0 + i \tilde v_0 \eta_\br) {\rm e}^{- i \tilde\lambda_0 t} \right. \nonumber\\
&& \phantom{\sqrt{n_0}} + \left. \beta (\tilde v_0 - i \tilde u_0 \eta_\br) {\rm e}^{i \tilde\lambda_0 t} \right],
\eea
where $\tilde\lambda_0$ is the lowest energy eigenvalue of the post-quench Hamiltonian, and
\bea
\alpha &=& (u_0 \tilde u_0 + v_0 \tilde v_0) \\
\beta &=& (u_0 \tilde v_0 - v_0 \tilde u_0).
\eea
This leads to a time dependent density
\bea
n(\br,t) &=& |\Psi(\br,t)|^2 \\
&=& n_0 \left[1 - 2 \alpha\beta \eta_\br \sin(2|\tilde\lambda_0| t)\right],
\eea
which exhibits staggered modulations on top of the uniform background average with a frequency $2|\tilde\lambda_0|$ and an amplitude, which depends on the degree of the quench. For a weak quench, where $\tilde u_0,\tilde v_0$ are close to $u_0,v_0$, the amplitude is small; however the amplitude can be significant for a strong quench.

\subsection{Approximate analysis for interacting bosons in the absence of a trap}
Let us now consider the effect of interactions on a weak quench in the absence of a trap, where $J_x$ suddenly decreases from 
$J$ to $J + \delta J < J$ at time $t=0$ while keeping $J_y=J$ constant. (Note that since we start from the isotropic case, and 
since we are studying currents and densities, which are both gauge invariant quantities, we would get exactly the same results for a quench along the $y$-direction.) Because the quench conserves crystal momentum in the reduced Brillouin zone, we can write the post-quench wavefunction in the form $\Psi(\br,t) = A(t) + B(t) \eta_\br$, where $A(t)$ and $B(t)$ denote time dependent complex coefficients of the uniform and staggered component of the wavefunction in real space. The full time-dependent GPE then reduces 
to a pair of non-linear ordinary differential equations for $A(t)$ and $B(t)$, given by
\bea
\!\!\!\! i \frac{d A}{d t} &\!\!=\!\!& \tilde\epsilon_0 A - i\gamma_0 B + U (A|A|^2 + 2 A |B|^2 + B^2 A^*), \\
\!\!\!\! i \frac{d B}{d t} &\!\!=\!\!& i \gamma_0 A - \tilde\epsilon_0 B + U (B|B|^2 + 2 B |A|^2 + A^2 B^*),
\eea
where $\gamma_0 = - 2 J \sin(\phi/2)$ and $\tilde\epsilon_0 = - 2 (J + \delta J) - 2 J \cos(\phi/2)$. It is easy to check that the total density does not change, since $d/dt (|A|^2 + |B|^2)=0$. However, as discussed in detail in Appendix C, the staggered differential density $\Delta n_{AB}(t) = 2 (A^* B + B^* A)$ can be shown to approximately obey the simple harmonic
equation
\bea       \label{freq}
\frac{d^2 \Delta n_{AB}}{d t^2} \approx - \Omega^2 \Delta n_{AB},
\eea
where 
$                                    
\Omega^2 = 4 [\tilde\lambda_0^2 + U n_0 (\tilde{\epsilon}_0 \epsilon_0 + \tilde{\gamma}_0 \gamma_0)/ |\lambda_0|].
$
Using the initial condition on the equilibrium currents, the solution to this can be written in the intuitive form
\be   \label{densitydynamics}
\Delta n_{AB}(t) =  4 \left(\frac{\delta J}{J}\right)  \frac{{\cal I}}{\Omega} \sin(\Omega t)
\ee
where ${\cal I}$ is the initial current on each bond, given by
\be
{\cal I} = \frac{4 J^2 n_0 \sin(\phi/2)}{|\lambda_0|}\,,
\ee 
$\delta J /J $ represents the fractional change in the hopping along the $x$-direction (which is also the instantaneous fractional change in the current along the $x$-direction induced by the quench), and the factor-of-four in the front arises from two $x$-bonds having been weakened by the quench. These sublattice density oscillations thus directly reflect the presence of staggered currents in the initial state; the amplitude of these oscillations depends linearly on $\delta J$ for a weak quench, while the frequency increases with increasing interaction strength.
Knowing $J$ and $\delta J$, a measurement of $\Delta n_{AB}(t)$ and its oscillation frequency $\Omega$ would thus 
provide quantitative information about the initial equilibrium current ${\cal I}$, which can be compared with the 
theoretically expected value quoted above. 

In case experimental imperfections are too strong to observe oscillations with a well-resolved frequency, the initial build-up of the density pattern might be used to extract ${\cal I}$. While this approach is unaffected by e.g. spatial variations of $\Omega$, it rests
on the ability to detect small changes in the sublattice populations.

\subsection{Numerical study of quench dynamics in the presence of a trap} 
\begin{figure}[t]
	\includegraphics[width=0.48 \textwidth]{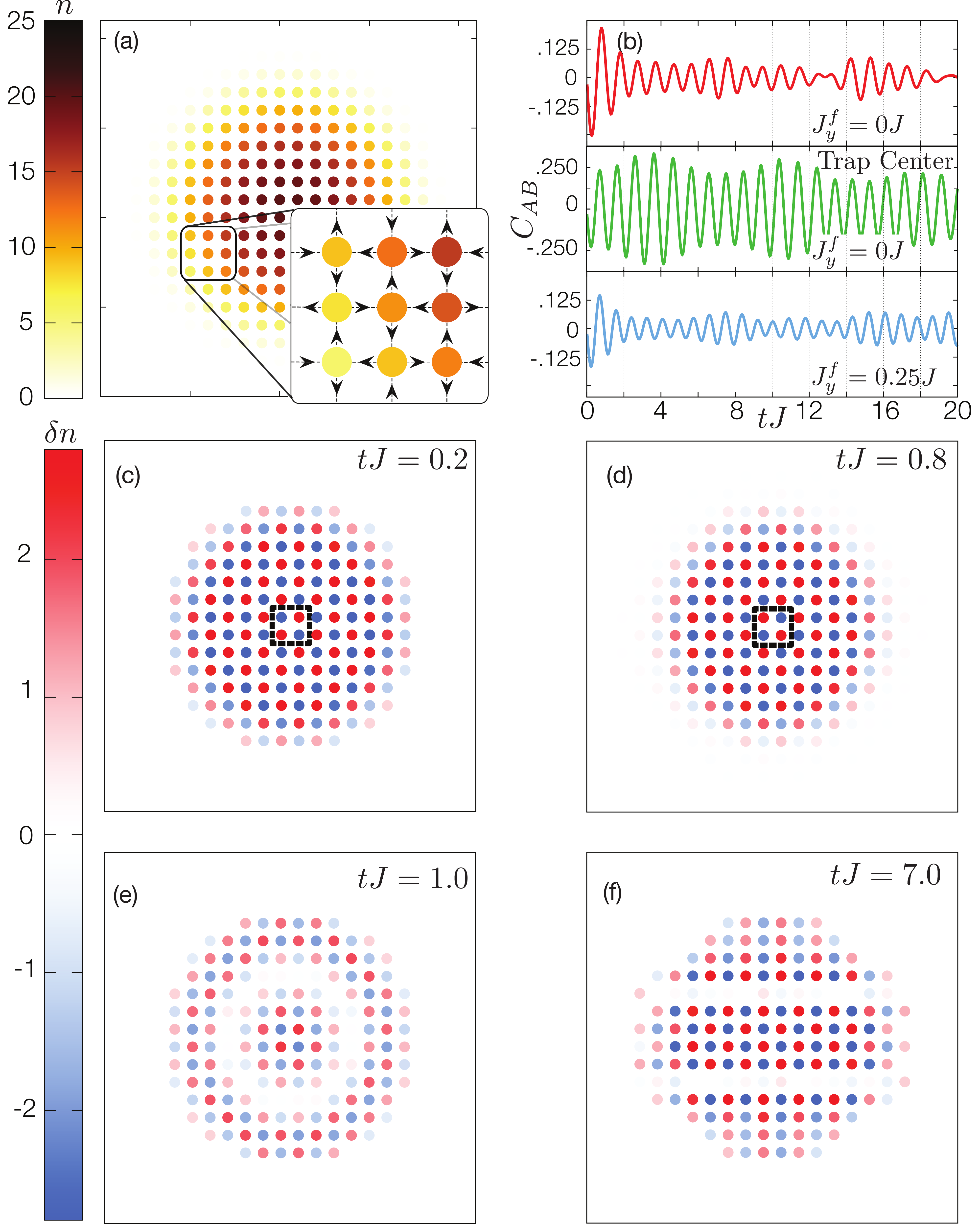}
	\caption {(Color online) Density pattern of 2D Bose superfluid in a staggered flux gauge field following a quench. (a). Initial density profile and (inset) current pattern of condensate ground state at half-filling.   (b). Time dependence of the sublattice 
density contrast $C_{AB} (t)$ for a staggered flux of $|\phi|=\pi/2$ and $U=0.2 J$ for different cases: (i) Trap average
of $C_{AB} (t)$ following 
a complete quench from $J^i_x=J$ to $J^f_x=0$ (top, red), (ii) Average of $C_{AB} (t)$ over a central $4\times 4$ region
(centre, green) following 
a quench from $J^i_x=J$ to $J^f_x=0$, and (iii) Trap average
of $C_{AB} (t)$ following 
a partial quench from $J^i_x=J$ to $J^f_x=0.25$ (bottom, blue).  (c -- f), Change in local density, $\delta n$ 
(relative to original density), at different times following a quench $J^i_x=J \to J^f_x=0J$ with 
$U=0.2J$. The marked region in (c) and (d) indicates the central four-site plaquette.}
\label{Fig:boson_stag}
\end{figure}

Having understood the underlying quench dynamics of the staggered flux state in the bulk, we now reintroduce a harmonic trap potential. The equilibrium state is solved self-consistently and leads to a superfluid ground state with staggered loop currents as shown in Fig.~\ref{Fig:boson_stag}(a) for a system with linear length $L=22$, $V_0=0.07J$, $U=0.2J$ and an average filling factor of $n_0=4$. The smooth density profile of the ground state reflects the trap potential, but it does {\it not} reveal the currents induced by the gauge field.

Starting with the equilibrium ground state, we perform a quench along the the $x$-direction and study the subsequent density dynamics. As directly seen in Fig.~\ref{Fig:boson_stag}(c,d), the condensate develops striking checkerboard oscillations at 
$t > 0$ that reflect the underlying current order. These oscillations can be monitored by the contrast of the spatial sublattice 
density modulations $C_{AB} = [N_A(t) - N_B(t)]/N$ shown in Fig.~\ref{Fig:boson_stag}(b) (with $N$ being the total number of
bosons). Information about the direction of circulation on a plaquette is easily discerned from the density pattern established after a short time period; since the quench is in $J_x$, the initial build up of density is on sites that have currents flowing into them along the strong $J$-bonds oriented along the $y$-direction. After a short time has passed, the density buildup reaches a maximum and the flow is reversed, resulting in ``plasma oscillations'' between the two checkerboard patterns. The frequency of these oscillations scales as $\sim \sqrt{ (\tilde\lambda_0^2 + U n_0 (\tilde{\epsilon}_0 \epsilon_0 + \tilde{\gamma}_0 \gamma_0)/ |\lambda_0|))}$ as shown earlier; it thus varies slowly with position due to the inhomogeneity of the density in the trap.

In order to compare the numerically computed dynamics with the analytical results of the previous section, we compute the local sublattice density contrast 
$C_{AB}^{(c)} = \langle n_A(t) - n_B(t) \rangle_c$, over a region of $4 \times 4$ sites in the center of the trap. 
We find that $C_{AB}^{(c)}(t)$ oscillates with a significant amplitude (up to $\sim 25\%$ contrast) and it can be fitted with the form in Eq.~(\ref{densitydynamics}).  The value of the current extracted from such a fit is $\sim 14.8J$, which is remarkably close to the value obtained from the analytical expression 
$4 n J\cos(\phi/8)\sin(\phi/8) \simeq 14.9J$ (the parameters are the central density $n_0^{(c)} = 19.4$ and  $\phi = \pi/2$).
The extracted value of the oscillation frequency, $\Omega = 6.72 J^{-1}$, also agrees well with the analytic result of $6.68 J^{-1}$ given by Eq.~(\ref{freq}). It should be noted that when the system is taken altogether, the early-time dynamics would provide a better fit to the average current, which is dominated by the central region.

The sublattice density difference when integrated over the entire trap exhibits some degree of dephasing and damping due to the density inhomogeneity; nevertheless, given our finite system size, the sublattice density oscillations persist out to fairly long times, $t J \gg 1$, as seen in Fig.~\ref{Fig:boson_stag}(b). At these times, 
we find additional long-wavelength modulations superimposed on the checkerboard density pattern (see Fig.~\ref{Fig:boson_stag}(e,f)). Prominent spherical density waves emanate periodically outward from the centre, which we attribute to the spatial variation of the ``plasma frequency'' resulting from the radial variation of the density $|\Psi_\br|^2$ in the trap. Furthermore, the cloud shape shows oscillatory distortions into an ellipse due to the anisotropy of the final tunneling $J^f_x < J_y$.

\subsection{Effect of random noise in the initial state} 

As a simple test of the robustness of the quench procedure to condensate depletion, we compute the quench dynamics of a state with imposed random phase fluctuations so as to mimic thermal fluctuation effects. We first add a small random and uncorrelated phase shift to each site, which is chosen in the range $(0,\delta\theta_{\rm max})$. This `random' state is evolved for a long period of time according to the (unquenched) GP equation to allow the system to equilibrate into a viable `thermal
state', which now supports correlated phase and density fluctuations. Starting from this thermal state, we next perform the sudden quench by evolving this state according to the \emph{quenched} GP equation, as before, and analyze its dynamics. 
We consider two cases, one with small fluctuations
$\delta\theta_{\rm max}^{(1)}=0.5$rad and one with moderate fluctuations $\delta\theta^{(2)}_{\rm max}=1.0$rad. To the extent that these fluctuations lead to states that mimic typical states from a thermal ensemble, 
both realizations of phase fluctuations result in superfluid states well below the 
Berezinskii-Kosterlitz-Thouless transition; details are given in Appendix D.

\begin{figure}
\includegraphics[width=0.49	 \textwidth]{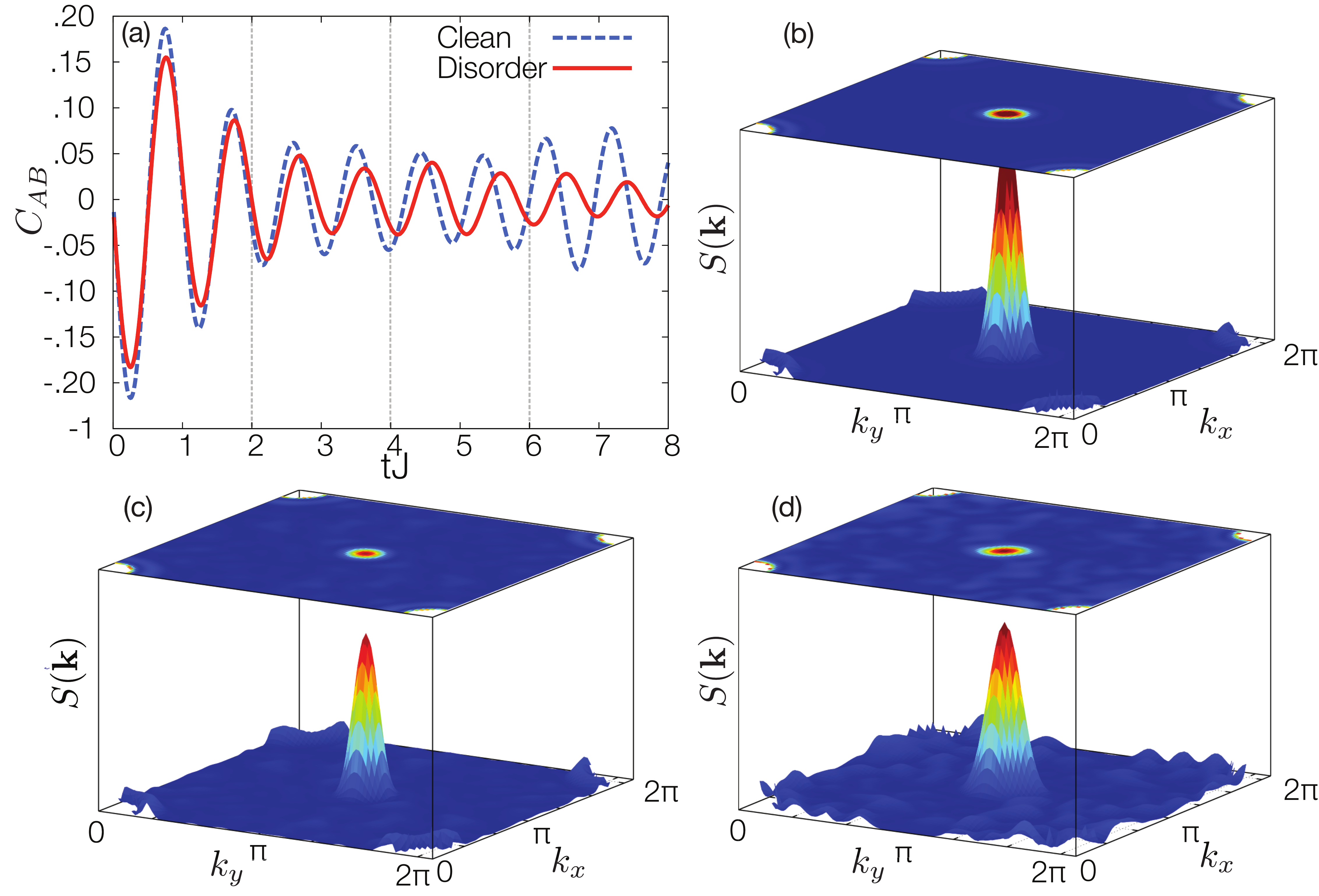}
\caption{(Color online) Dynamical density pattern of interacting bosons on a 2D square lattice in a staggered flux gauge field following a quench in the presence of noise. (a) Comparison of the time dependence of the sublattice density contrast $C_{AB} (t)$ for a staggered flux of $\phi=\pi/2$ following a quench of $J^i_x=J$ to $J^f_x=0$ without noise (solid red) and with noise (dashed blue).  Noise is incorporated by including a random local phase fluctuation in the range $(0, \delta\theta_{max})$ with $\delta\theta_{max}=0.5$ at each site, and evolving the system according the GPE for a long period of time before implementing the quench. Normalized structure factor at $tJ=0.25$ for a system (b) without noise and (c,d) with the presence of noise.  Noise is incorporated by adding a random phase, with fluctuation magnitude (c) $\delta\theta^{(1)}_{max}=0.5$ and (d) $\delta\theta^{(2)}_{max}=1.0$, locally to each site and allowing the system to equilibrate. Note, the momentum peak close to $(0,0)$ associated with the average density distribution has been removed for clarity and the axes in (b -- d) are equal.}
\label{Fig:boson_noise}
\end{figure}
                                                  
As shown in Fig.~\ref{Fig:boson_noise}(a), the initial sublattice density contrast in a given state with noise, or averaged over several such realizations, is similar to that of the clean system; however, whereas the oscillations persist for a long time in the ground state, the oscillations in such a`thermal state becomes incoherent after a few  periods. Furthermore, even at short timescales, we find that: (i) the amplitudes are no longer equal to that of the ground state quench, and (ii) the oscillation frequency is slightly shifted compared to its zero temperature result.

A further impact of thermal fluctuations is found in the broadening of the spectral peak in the structure factor shown in Fig.~\ref{Fig:boson_noise}(b-d).  (Note, that the strong momentum peaks about $(0,0)$ have been removed for clarity.)  The large single spectral peak at $(\pi,\pi)$ in the clean system, shown in Fig.~\ref{Fig:boson_noise}(b), is replaced by a broader peak in the noisy systems shown in Fig.~\ref{Fig:boson_noise} (c,d), where $\delta\theta_{max}$ is equal to $0.5$ rad and $1.0$ rad in (c) and  (d), respectively. Despite the fluctuations and broadening, the $(\pi,\pi)$ peak at early times can still easily be discerned above the
background. We observed that the real-space density pattern also displays a discernible checkerboard-like tendency even in the presence of moderate thermal noise.

\section{Bose superfluid in a stripe synthetic flux background}

\begin{figure}[t] 
	\includegraphics[width=0.48 \textwidth]{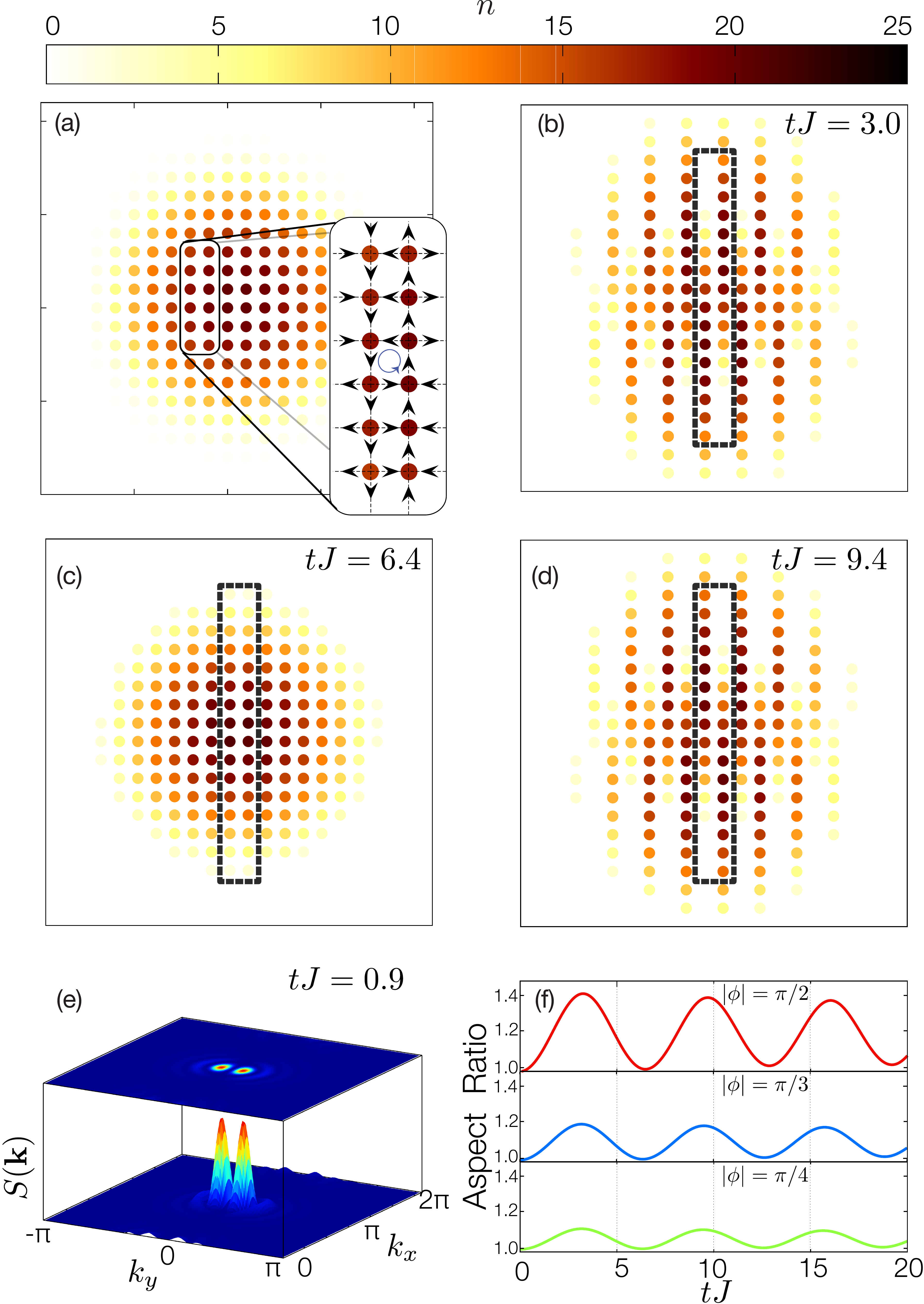} \caption {(Color online) Dynamical
density pattern of interacting bosons on a 2D square lattice in a stripe-like magnetic flux pattern
following a quench.  (a) Initial density profile and (inset) current pattern of condensate ground
state at half-filling. ( b -- d): Change in local density, $n$, at different times following a quench $J^i_x\!=\! J \to J^f_x\! =\! 0$ with $U\! =\! 0.2J$.  
The circled region indicates the central elongated vortex with a nonzero quadrupole moment.  
(e) Density structure factor normalized to the peak maximum and with the momentum peak at $(0,0)$ associated with the average density removed for clarity.  
(f) Aspect ratio of the cloud as function of time for various fluxes.}
\label{Figboson_stripe} \end{figure}

We next consider the Bose-Hubbard model in the presence of a striped magnetic flux pattern as realized in Ref.~\cite{bloch.prl2011} (see Fig.~\ref{Fig:quench_schematics}(b)). We choose $J_{\br,\br+\hat{y}} \! = \! J_y$ and $J_{\br,\br + \hat{x}} \! = \! J_x \exp [i (-1)^{x} \phi y]$, so that we enclose fluxes $\pm \phi$ through each plaquette that lies along a stripe in the $y$-direction, and solve for the equilibrium ground state by minimizing the GP energy functional for $J_x=J_y=J$, and for weak interactions $U = 0.2 J$, with $L=22$, $V_0=0.07J$, and an average filling $n_0=4$. We find a superfluid with vertically striped loop currents depicted in Fig.~\ref{Figboson_stripe}(a), which resembles a stripe pattern of `long vortices' that are highly elongated along the $y$-direction. Again, the smooth equilibrium density pattern is reflective of the underlying trap potential but it reveals {\it no} information about the underlying currents.

Upon quenching $J_x$, the superfluid generates a {\it density} pattern that strikingly reflects the underlying equilibrium striped currents. Each vertically elongated loop
forms four quadrants of alternating high and low density, giving rise to an oscillatory quadrupole moments as can be seen in Fig.\ \ref{Figboson_stripe}(b--d).
Evidence for striped density pattern can also be found in the structure factor shown in Figure \ref{Figboson_stripe}(e).  The small momentum modes about $(0,0)$ that are attributed to the average density are subtracted off for clarity, but we emphasize that it is not necessary to know the original density distribution before the quench for any of the analysis. The structure factor shows two dominant spectral peaks in proximity to $(\pi, 0)$ but slightly shifted by ${\bf q}=(0,\pm q)$.   This small momentum shift is a reflection of the additional long wavelength component that originates from the anti-nodal line in the density pattern running along the centers of the elongated quadruples at $y=0$.  Since this $q$ scales as $1/L$, the two peaks will merge toward $(\pi,0)$ for larger system sizes.    

Quenching $J_y$ rather than $J_x$ leads to similar early-time density patterns; however, the oscillatory dynamics that occur are much faster.  In addition, the density oscillations in this case rapidly
decohere when compared with the quench in $J_x$. 
These observations hint that a very different mechanism is at play for the two quenches. 

To better understand the dynamics, it useful to consider the infinite system without a trap.  The initial density pattern can be shown to be uniform with alternating stripe currents flowing up and down along the bonds that lie along the $y$-direction, which has a magnitude of $2J_y \sin(\phi/2)$ for $\phi<\pi/4$.  In this limit, neither quench (in $J_x$ or $J_y$) leads to any density modulation and no information pertaining to the current can be obtained from such quenches. This situation occurs for any system with constant current along a given bond direction. Despite this, information about the current can still be inscribed onto the density \emph{when the quench is performed in the presence of a trap}. 

Consider again the stripe flux state in the presence of a trap. Close to the center of the trap the current again alternates direction and flows along the $J_y$ bonds, however, there are now strong edge currents in the vicinity of the cloud boundary (in the above simulation, the edge currents die off as the center of the trap is approached, but are still present due to finite size affects).  When the $J_x$ hopping is quenched, the bosons now continue to flow in the direction of current and, instead of leaking into the edge current, will flow up the trap potential.  This generates density oscillations along the vertical chains whose frequency is determined by the trap profile and the effective mass of the bosons. Since, the initial direction of the current flow alternates along each of the vertical chains, the phase difference between the density oscillations of neighbouring chains is $\pi$, thereby producing the stripe pattern. In contrast, a quench in $J_y$ directly probes the edge currents in the system and, although the density pattern is again striped, the oscillation frequency is not determined by the trap potential.

The least demanding experiment is a measurement of the aspect ratio of the cloud, $\sqrt{D_{x^2}(t)/D_{y^2}(t)}$ 
where $D_{x^2\,(y^2)}(t)=\sum_\br n(\br,t)x^2\,(y^2)$, as a function of time
(see Fig.~\ref{Figboson_stripe}(f)). Notice that only the oscillation {\it amplitude} is discernibly affected by the value of the magnitude of the flux per plaquette, while the oscillation frequency is essentially governed by the trap frequency and thus is practically independent of flux. The variation of the amplitude reflects the differences in the initial currents for different flux values.

\section{Triangular lattice frustrated Bose superfluid}
\begin{figure}
\includegraphics[width=3.4in]{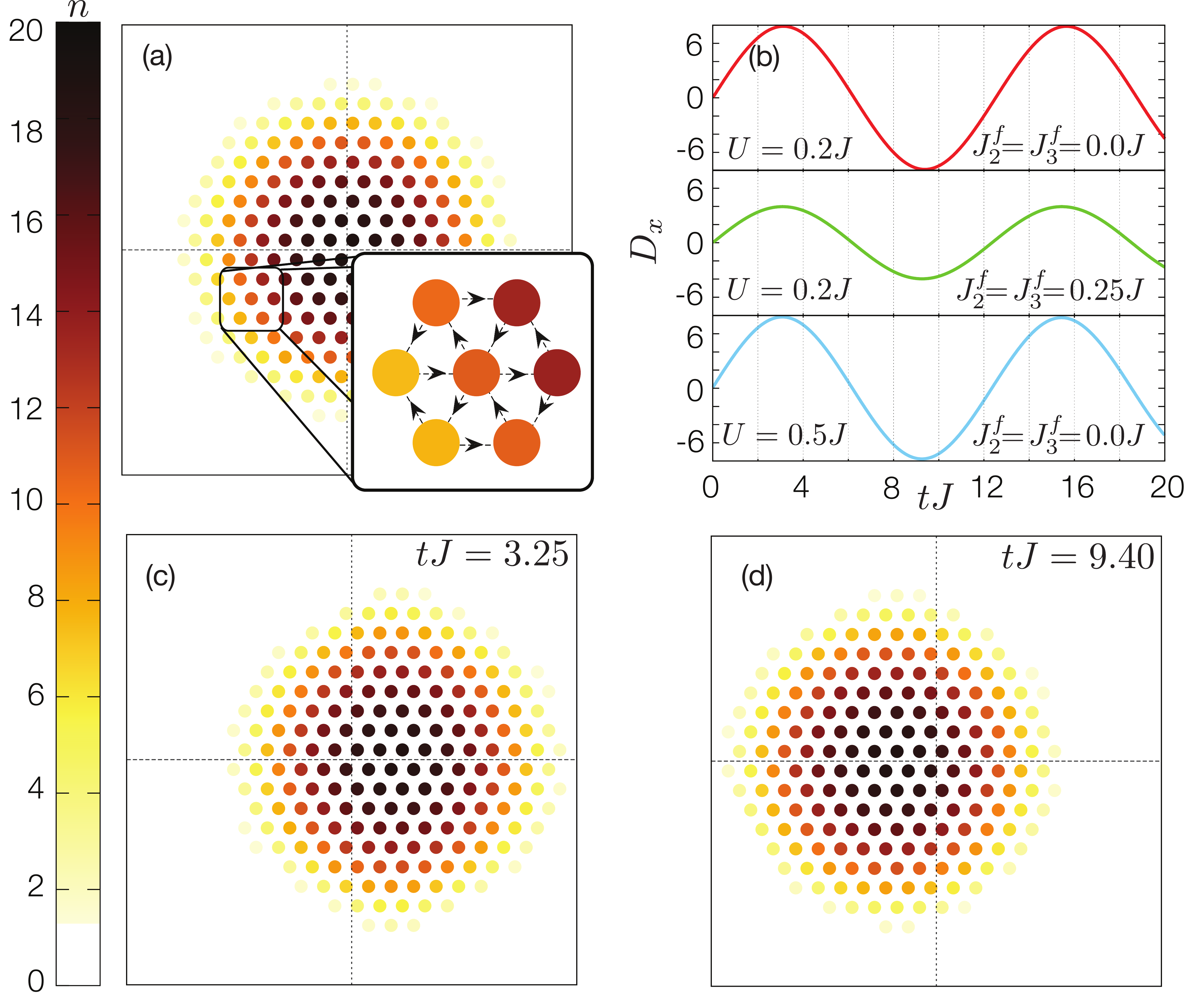} 
\caption{(Color online) Density dynamics for spinless 2D bosons in a staggered flux gauge field on a
triangular lattice after simultaneously quenching the $J_2$ and $J_3$ bonds.  (a) Initial density pattern and (inset) current pattern at half filing with flux $|\phi|=\pi/2$.  (b)  The dipole moment as a function of time for various parameters. Distances are measured in units of the lattice constant. (c,d) Density profile at the indicated times showing the displacement of the cloud.  The crosshairs indicate the center of the trap.} 
\label{Fig:Tri}
\end{figure}

A similar situation arises for a triangular lattice in the presence of a staggered flux state (see Fig.~\ref{Fig:quench_schematics}(c)).  Consider the initial state to be the ground state of a system with $\phi=\pi/2$ flux per plaquette and further $J_1=J_2=J_3=J$, where $J_1$, $J_2,$ and $J_3$ are the magnitudes of the tunnel couplings along the $(1,0)$, $(1/2,\sqrt(3)/2)$ and $(-1/2,\sqrt(3)/2)$ bond directions, respectively, and $U=0.2J$.  As shown in the inset of Fig.~\ref{Fig:Tri}(a), the ground state currents flow along the bond directions, as they do for the infinite system without a trap.  Hence, we again rely on the combined effect of the quench with the trapping potential to transcribe information about current flow onto the density profile.

Since there are now three unique bond directions, it is necessary to quench the hopping along two of the bond directions simultaneously, which we choose to be $J_2$ and $J_3$.  Here, the dynamics is very similar to that of the $J_x$ quench for the stripe flux considered previously.  Each chain decouples and the initial current causes the density to oscillate in the trap potential.  This time, however, the direction of current flow is the same along neighbouring chains, resulting in a uniform oscillation of the entire cloud along the unquenched bond direction. (Note that unlike the checkerboard
case, this current pattern is at ${\bf q}=0$, so that the quench does not produce any density modulations with nonzero
Fourier components.)

This oscillation is best observed by monitoring the time dependence of the dipole moment $D_x= \sum_{\vec r} n(\vec r,t) x$, as shown in Fig.~\ref{Fig:Tri}(b);
a larger equilibrium current in the initial state will lead to a 
larger \emph{amplitude} for such center of mass oscillations following a quench.
Notice, however, that  the oscillation frequency is nearly independent of the magnitude of the quench.
This substantiates the idea 
that the oscillation frequency set by the trap stiffness and is independent of the initial current. Furthermore,
the oscillation frequency is independent of the interaction strength, as we confirm numerically in Fig.~\ref{Fig:Tri}(b), since it involves only center-of-mass oscillations. One should keep in mind that the oscillation frequency can shift if the initial flux per plaquette is changed.  This is expected because the bosons will initially condense into a state with different crystal momentum and lattice effects will change the effective mass of the condensate.

\section{Spinless fermions in a staggered magnetic flux}
Motivated by our study of quench-induced density dynamics for bosons, we next turn to noninteracting fermions 
in a staggered flux background \cite{stagflux1}. We study the Hamiltonian $H_{\rm sf}  = - \sum_{\br,\br'} J^\pdg_{\br,\br'} f^\dg_\br f^\pdg_{\br'}$, where $J_{\br,\br+\hat{x}} \! = \! J$ and $J_{\br,\br + \hat{y}} \! = \! J_y \exp [i (-1)^{x+y} \phi/2]$, leading to staggered checkerboard fluxes $\pm \phi$ \cite{hemmerich.stagsf}. To make analytical progress, we ignore the harmonic trap in the discussion below. As we have seen previously for bosons, and as discussed below, the trap does not qualitatively affect our conclusions, and we can also directly apply our results to the central region of the trapped gas. In momentum space, the Hamiltonian takes the form
\bea
H_{\rm sf}  = {\sum_{\bk}}^{\prime} \Omega_\bk
\Psi^\dg_\bk
(\cos\theta_\bk \tau^z + \sin\theta_\bk \tau^y)
\Psi^\pdg_\bk,
\eea
where $\bQ\equiv (\pi,\pi)$, $\Psi^\dg_\bk = (f^\dg_\bk, f^\dg_{\bk+\bQ})$, and $\tau^{y,z}$ are Pauli matrices. The prime on the momentum sum implies that only momenta in the reduced Brillouin zone are included. Here, we have defined
$\Omega_\bk \! = \! \sqrt{\varepsilon^2_\bk + \gamma^2_\bk}$, $\cos\theta_\bk \! = \! 
\varepsilon_\bk/\Omega_\bk$,
and $\sin\theta_\bk \! = \! \gamma_\bk/\Omega_\bk$, with
\bea
\varepsilon_\bk & = & - 2 (J \cos k_x + J_y \cos \frac{\phi}{2} \cos k_y) \\
\gamma_\bk & = &  - 2 J_y \sin \frac{\phi}{2} \cos k_y.
\eea 
This leads to mode energies $\pm \Omega_\bk$ in the initial state.

\begin{figure}[t]
\includegraphics[width=3in,height=2in]{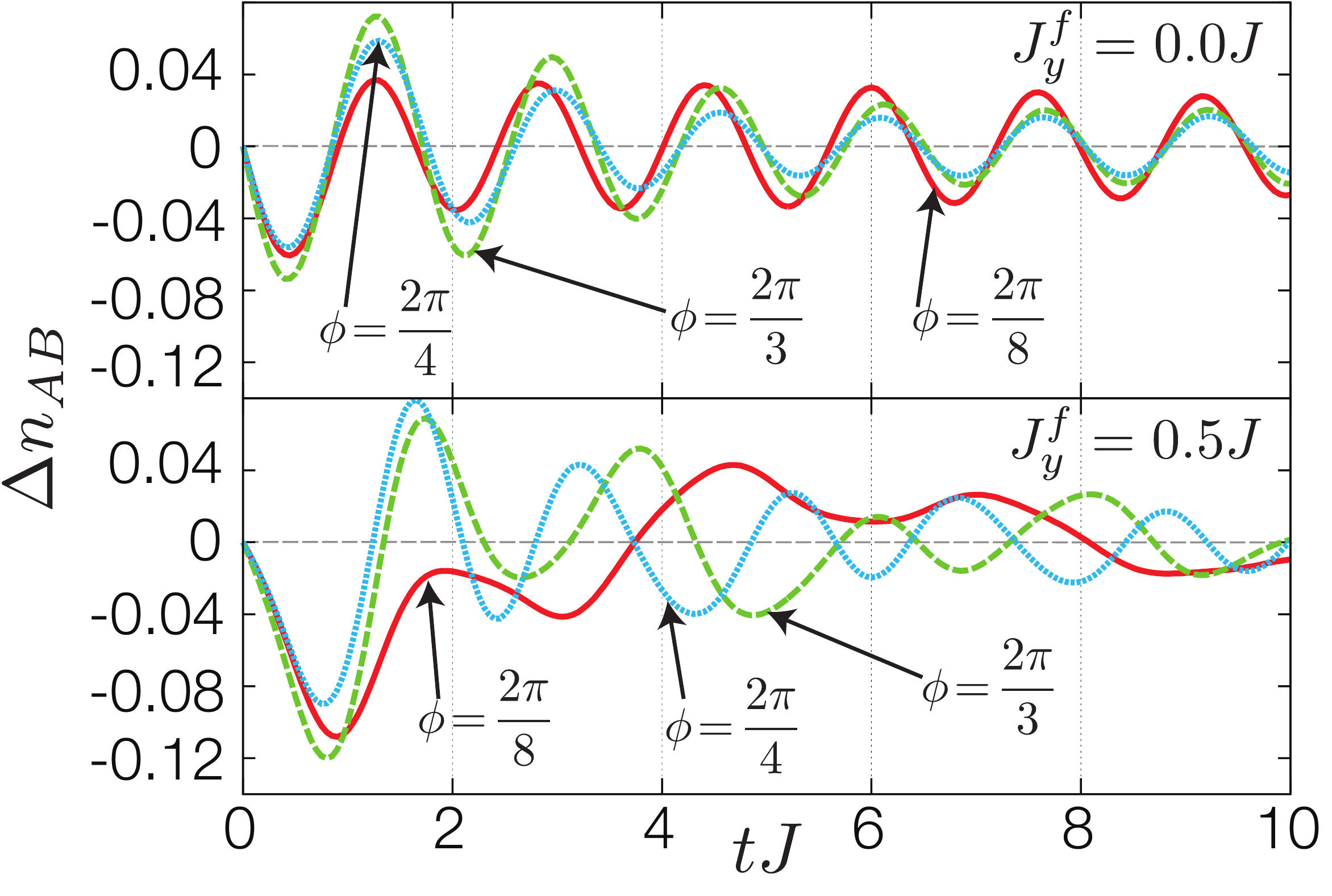}
\caption{(Color online) Time dependence of the sublattice density difference 
$\Delta n_{AB} (t)$, for noninteracting spinless fermions on a 2D square lattice at a
filling of $n_0 = 0.4$, flux $\phi=\pi/2$,
following a
quench from (upper panel) $J^i_x=J \to J^f_x=0J$ or (lower panel) $J^i_x=J \to J^f_x=0.5J$.}
\label{Fig:fermion_free}
\end{figure}

Imagine fermions initially filled into negative energy states $-\Omega_\bk$ up to a Fermi energy $E_F$, and then quenching $J_x$ from $J_x^i \! \to \! J_x^f$ at time $t\! =\! 0$. Such a translationally invariant quench ensures that different momentum pairs $(\bk,\bk+\bQ)$ stay decoupled from each other. Nevertheless, this quench
instantaneously changes 
$\varepsilon_\bk \to \tilde{\varepsilon}_\bk$, and 
$\gamma_\bk \to \tilde{\gamma}_\bk$, so that we modify
$(\Omega_\bk,\theta_\bk) \to (\tilde\Omega_\bk,\tilde\theta_\bk)$.

This means that while the initial quasiparticle occupation numbers are set by the initial dispersions and the chemical potential, the subsequent dynamics is then determined by the final Hamiltonian. Since the final Hamiltonian is also translationally invariant, the various momentum states stay decoupled after the quench,
but undergo the analogue of ``spin precession'' in the two-level $(\bk,\bk+\bQ)$ space. 

To compute the density modulation between the two sublattices at a subsequent time, 
$\Delta n_{AB} (t) \equiv  (n_A-n_B)$, we write
\bea
\Delta n_{AB} (t) &=& \frac{2}{M} \sum_\br (-1)^{x+y} \la f^\dg_\br f^\pdg_\br \ra_t \nonumber\\
&=& \frac{2}{M}
{\sum_\bk}^{\prime} (\la f^\dg_\bk f^\pdg_{\bk+\bQ} \ra_t + \la f^\dg_{\bk+\bQ} f^\pdg_\bk  \ra_t)
\eea
where $M$ is the number of lattice sites.
Carrying out the algebra, details of which are given in Appendix E, we find
\be
\Delta n_{AB} (t) = \frac{2}{M} {\sum_{\{\bk\}_{\rm occ}}}^{\!\!\!\prime} \sin(\theta_\bk - \tilde\theta_\bk) \sin(2 \tilde\Omega_\bk t)
\ee
where the momentum sum runs only over {\it initially occupied} states in the reduced Brillouin zone.

A numerical evaluation of the sum allows us to plot the sublattice density oscillations, shown in Fig.~\ref{Fig:fermion_free} for $J_x^i \! = \! J$, $J_x^f \! =\! 0.0 J$ and $J_x^f \! =\! 0.5 J$, a fermion density of $n_0 \! = \! 0.4$ per site, and various staggered flux values.
These oscillations exhibit multiple frequencies due to the large number of occupied fermion modes. However, over the entire range of displayed fluxes, and a wide range of densities $n_0 \! \sim \! 0.3$ -- $0.5$ near half-filling, we find that the dominant oscillation frequency arises from initially occupied states near $\bk\! =\! (0,\pi)$ due to a van Hove singularity in the density of states. Picking this single mode $\bk=(0,\pi)$ in the above momentum sum leads to an estimated, nearly density-independent, dominant oscillation frequency $(2 \tilde\Omega^*) \approx  4 \sqrt{J^2+(J^f_x)^2 - 2 J J^f_x \cos\frac{\phi}{2}}$. Both, the flux dependence of this oscillation frequency for a partial quench, and its flux independence for a complete quench with $J^f_x=0$, are in quantitative agreement with the numerical data in Fig.~\ref{Fig:fermion_free}. The larger density of states near $\bk \! = \! (0,\pi)$ also enhances the signal amplitude for fillings close to $n_0 \! = \! 1/2$. The weak density dependence of $\Delta n_{AB}(t)$ over a range of fillings indicates that trap induced inhomogeneities will not significantly affect these oscillations.

\section{Topological states with edge currents}
Finally, we turn to gapped yet
topologically nontrivial states such as quantum Hall insulators, Chern band insulators, 
or quantum spin Hall insulators, all of which have bulk gaps but
support topologically protected {\it edge} currents.
At a fundamental level, Chern band insulators are no different from integer quantum Hall states in the
Hofstadter model, as discussed in recent work \cite{flatband}.  Both systems
involve fluxes threading through plaquettes of the lattice with no net flux
over an appropriately defined unit cell; in the Hofstadter model, this unit
cell is the magnetic unit cell.

 Proposals to obtain such fluxes in experiments on cold atoms exist in the literature
\cite{cooper,trotzky,TI, TI3D}. The simplest models of 2D quantum spin Hall states or topological
insulators, such as the Kane-Mele model \cite{kanemele.2005}, may be viewed as two independent copies of
quantum Hall insulators or Chern band insulators, with the two copies being labelled by a well-defined
spin quantum number (equivalently `hyperfine state' for an atom) and experiencing opposite magnetic
fluxes. Much of the physics we discuss below, which involves studying the density dynamics following a
quantum quench, will then be applicable to such quantum spin Hall states if one can experimentally probe
the density of each spin species.

While recent work has focused on extracting
the nontrivial band topology from time-of-flight measurements \cite{zhaospielman,pricecooper} 
or spectroscopy of the edge modes \cite{gerbier,gerbier2},
here we explore
density dynamics induced by the unidirectional quench for lattice fermions in a uniform
magnetic field.
For concreteness and reasons of simplicity, we
consider fermions on a 2D square lattice
with a uniform magnetic flux $\phi\! \! = \! \! 2\pi/3$ per plaquette.  Similar uniform flux configurations have recently been established in cold atom experiments by rotating the optical lattice \cite{rotating1,rotating2}.
The resulting particle-hole symmetric Hofstadter
spectrum \cite{hofstadter} has three non-overlapping bands, with
Chern numbers
$+1,-2,+1$, so that `band
insulators' with {\it some} bands being completely filled support a nonzero quantized 
Hall conductance, and chiral 
edge currents, yielding lattice versions of integer quantum Hall (QH) states
in the continuum  \cite{tknn}.

\begin{figure}
\includegraphics[width=3.4in]{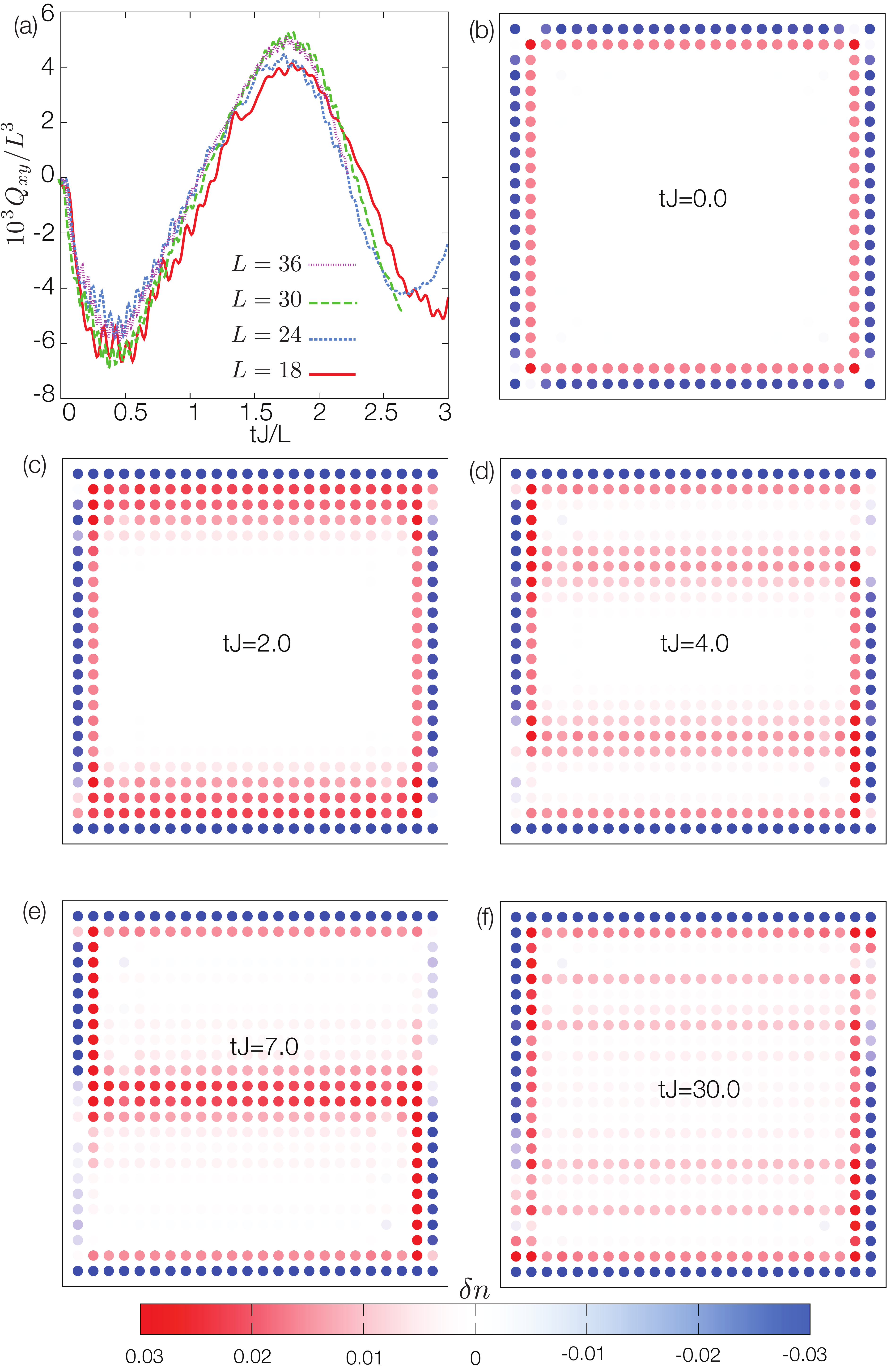}
\caption{(Color online) Density dynamics for
spinless fermions in the lowest Hofstadter band with flux $\phi=2\pi/3$
per plaquette on a square lattice following a quench
from $J_x^i=J$ to $J_x^f = 0.0 J$. (a) The scaled 
quadrupole moment 
$Q_{xy}(t)/L^3=(1/L^3) \sum_{\vec r} x y n(\vec r,t)$
for various system sizes versus the scaled time $t J/L$.
(b -- f) Stripe-like density modulations (for $L=24$, plotted as $n(\vec r,t) - n_0$)
moving from the $y$-edges into the initially incompressible bulk
at different indicated times.}
\label{Fig:QHE}
\end{figure}

We begin by numerically diagonalizing the Hamiltonian 
$H_{\rm QH}  = - \sum_{\br,\br'} J^\pdg_{\br,\br'} 
f^\dg_\br f^\pdg_{\br'}$ with $J_{\br,\br+\hx} = J  {\rm e}^{i \phi y}$ and $J_{\br,\br+\hy} = 
J_y$,
for $\phi=2\pi/3$,
with open boundary conditions on a $L \times L$ system,
and fill up the lowest band (and some edge modes) to get a
fermion filling $n_0 =1/3$.
We find that the ground state bulk
density is uniform (see Fig.~\ref{Fig:QHE} for $t=0$) and supports edge currents 
confined to an ``edge layer'' of thickness $\sim 2-3$ lattice sites where the
density also slightly deviates from its bulk value.
We next track the density 
dynamics following a quench from $J_x^i = J$ to  $J_x^f < J$, which is
easy to study once we compute the
initial and final spectrum and eigenstates. We note that the gauge choice
for the magnetic field (i.e., how exactly to include the vector potential) is
unimportant --- we could equally well have them along the $x$-bonds.

Viewing the chiral edge currents as analogous to that arising from a `vortex',
we expect the
quench to lead to quadrupolar density 
oscillations and current reversals, similar to what we found for the
`long vortex' in the stripe flux superfluid. 
Inspired by recent work on the
superfluid Hall effect of atomic bosons \cite{Leblanc},
we study the behavior of the 
quadrupole moment $Q_{xy}(t)= \sum_{\vec r} x y n(\vec r,t)$.
We find that $Q_{xy}(t)$ indeed displays oscillatory
sign reversals and, as seen in Fig.\ref{Fig:QHE}(a), the data for various $L$ 
collapse when
plotted as $Q_{xy}(t)/L^3$ versus $t/L$. The $t/L$ scaling shows that the 
oscillations occur due to transport across the system length $L$.
A simple scaling argument for an edge current induced oscillation shows
that $Q_{xy} \sim L^3$, as we also see numerically.
The numerical observations
are thus consistent with the quadrupolar oscillations being driven by edge currents.
For $\phi=-2\pi/3$, $Q_{xy}(t)$
has the opposite sign. Taken together, these observations provide strong
evidence that the initial state is a nontrivial insulator that is 
{\it incompressible in the bulk and supports 
chiral edge currents}.

In addition to this quadrupolar oscillation arising from edge modes, we find that 
the density in the bulk is also no longer constant following the quench. Instead,
quenching $J_x$ leads to $x$-oriented stripe patterns of the density, which 
originate at the edge and appear to propagate inward into the bulk. Such a
breakdown of the incompressible quantum Hall state in the bulk can be understood
physically by analogy to the physics of continuum Landau levels. A uniform
magnetic field in the continuum can be modelled in the Landau gauge
where we set $A_y=0$, $A_x= B y$ corresponding to a magnetic field $B \hat{z}$.
This leads to the Hamiltonian
\be
H_{LL} = \frac{1}{2 m_y} p_y^2 + \frac{1}{2m_x} (p_x + q B y)^2 + V(y)
\ee
where $q$ is the charge, and
for simplicity, we assume a confining potential $V(y)$ only along the $y$-direction 
and periodic boundary conditions along the $x$-direction. (In cold atom systems,
where the gauge field is produced artificially, only the product $q B$ is physical and tunable.)
In the absence of the confining potential, the eigenstates of this Hamiltonian take 
the form $\Psi_{n,k}(x) = {\rm e}^{i k x} \Phi_n(y-y_k)$ where $y_k= -k/q B$. Here $\Phi_n$
is the $n$th eigenstate of a harmonic oscillator with an energy $(n+1/2) \omega_L$,
with $\omega_L = q B/\sqrt{m_y m_x}$, and the particle in such a state is localized to 
the vicinity of $y=y_k$. If the confining potential is varying
slowly, with $\partial V/\partial y \ll \omega_L/\ell_B$ where $\ell_B = (\frac{m_y}{m_x})^{1/4} (1/\sqrt{q B})$
is the magnetic length, the eigenstates remain nearly unaffected by the confining potential
(up to a small shift of $y_k$) while the energy gets a correction $V(y_k)$. Imagine now
quenching the dynamics in the $x$-direction by sending the effective $m_x \to \infty$ suddenly.
This is analogous to sending 
$J_x \to 0$ on the lattice. The Hamiltonian
after the quench then takes the simple form
\be
H_{t > 0} = \frac{1}{2 m_y} p_y^2 + V(y),
\ee
which describes a free particle in a potential. Clearly $k$ remains a good quantum number.
This means that for each $k$, there are particles initially localized at different
points $y_k$ and described by an initial wavefunction $\Phi_n(y-y_k)$, which at time $t > 0$
are free to roll down the valley of this potential and delocalize (spread out). At time $t=0$, 
the particle density is uniform, so there are particles localized in the bulk starting at $y_k=0$ 
and going out all the way to the two edges at $y_k \sim \pm L/2$. For time $t > 0$, this
state evolves in time, and the density of the resulting state acquires modulations. This leads to
stripe modulations, with the net density depending on $y$ but not on $x$, with a time dependence
governed by a combination of the two effects above --- particles `rolling down' the potential and 
spreading of the initial harmonic oscillator wavepacket. This picture qualitatively accounts for 
the appearance of stripe modulations of the density in the bulk in the quenched state.
It also suggests that the stripe
modulation does not contribute
to the quadrupole moment dynamics, which is purely an edge current effect, consistent with the $L^3$
scaling in our numerical results.

It is natural to ask how the quench dynamics are modified when the boundary conditions are no longer of hard wall  type,
and the atoms are confined to a harmonic trap. In this case,
we continue to expect  bulk stripe-like density waves to emerge.  However, the density dynamics close to the `edges' that lie 
parallel to the quenched hopping direction (i.e.,\ the top and bottom sections of the trap) is expected to change.  
Instead of the density piling up at the boundary and reversing, the current will continue to flow up the trap potential
before eventually reversing. Hence, the frequency of the $Q_{xy}$ oscillation will be determined by the 
trap frequency and the effective mass of the fermions, and should match that of the bulk stripe density oscillations.  
Despite this, there will remain a definitive signature of edge state currents that can be found in the enhanced 
amplitude of the density modulations along the top and bottom edges.  (These edge states can be thought of as 
having a finite initial velocity perpendicular to the quenched hopping direction.)  This will lead to a \emph{shearing} of the cloud 
density and a finite $Q_{xy}$, which is expected to be easily discernible.

We expect Chern band insulators to exhibit similar quadrupolar density oscillations arising from the
currents at the edge. For quantum spin Hall insulators, with a conserved $S_z$ magnetization, 
we can imagine doing a similar quench experiment and measuring $Q^\upa_{xy} - Q^{\dna}_{xy}$, which would exhibit oscillations with a similar scaling. Such spin resolved quadrupole measurements
rely upon the recently demonstrated experimental ability to measure spin resolved densities
 \cite{weitenberg.nature2011}.
 
\section{Summary}

Atomic bosons and fermions in the presence of frustration or background synthetic gauge fields
carry mass currents with diverse current patterns or even form gapped topological phases with
edge currents. We have shown that anisotropic 
quantum quenches can yield a powerful probe of  such equilibrium current patterns 
of atoms in an optical lattice by converting them into measurable real-space density 
oscillations. In order to avoid exciting particles into the high energy bands of the periodic optical
potential,  the quench must be 
``adiabatic'' on time-scales comparable to the 
inverse interband gap, while also being ``sudden'' on time scales governing 
intraband dynamics. This requirement can be easily fulfilled in experiments since the
tunnel coupling between neighboring wells is exponentially suppressed when
the lattice depth is increased, while the energy separation between bands grows with the square-root
of the lattice depth \cite{bloch.rmp2008}.
Realizing our proposal for an experimental probe of currents would open up a new avenue to study exotic 
phases of ultracold atomic matter.

\acknowledgments

We thank Lindsay LeBlanc, Joseph Thywissen, Erhai Zhao, Xiaopeng Li, and Vincent Liu
for stimulating conversations. AP also acknowledges extremely useful discussions with Siddharth
Parameswaran, Anatoli Polkovnikov, Ludwig Mathey, Rafael Hipolito,
Gang Chen, and participants of the KITP workshop on
``Quantum Dynamics in Far from Equilibrium Thermally Isolated Systems''.
This research was funded by NSERC of Canada,
and supported in part by the National Science Foundation under Grant No. NSF PHY11-25915.

\section{Appendix}

\subsection{Numerical solution of equilibrium GP equation for trapped Bose superfluids}   
For Bose superfluids in the presence of a trap, the ground state is determined numerically using a
self-consistent minimization of the initial energy functional 
in Eq.~\ref{GPfunc} recast in terms of the local
mean-field density, 
\begin{eqnarray}\label{MF} H_{\rm MF}\left(n_\br\right) = - \sum_{\br, \br'}
J_{\br,\br'} \Psi^*_{\br} \Psi^\pdg_{\br'} + \sum_\br V_\br |\Psi_\br|^2 \notag \\ + U \sum_\br n_\br
\Psi^*_\br \Psi^\pdg_\br-\frac{U}{2} \sum_\br n_\br^2. \end{eqnarray} 
Here $\Psi_\br$ 
is the condensate wavefunction at lattice site $\br \equiv (x,y)$ and $n_\br=|\Psi_\br|^2$
is the average particle density at site $\br$, 
The self-consistent solution of Eqn.~\ref{MF} is one
where the density distribution computed using the eigenfunctions of $H_{\rm MF}$ equals the density
distribution $n_\br$ in the Hamiltonian.
In order to find the self-consistent solution of Eqn.~\ref{MF} corresponding to the initial pre-quench state, 
we follow a simple iterative procedure. We start with a trial density distribution $n^{(0)}_\br$ corresponding to a Thomas-Fermi profile and solve for the single particle ground state eigenfunction of $H_{\rm MF}\big(n^{(0)}_\br \big)$. The corresponding many-body condensate wave function is simply given by the normalized single particle solution with a multiplicative factor of $\sqrt{M}$, 
explicitly $\Psi^{(0)}_\br=\sqrt{M}\psi^{(0)}_\br$ (where $M$ is the number of lattice sites).  
The condensate density is then determined by $\widetilde{n}^{\,(0)}_\br=|\Psi^{(0)}_\br|^2$.  This is used to generate a new trial density distribution via the relation $n^{(1)}_\br=(1-\alpha)\widetilde{n}_\br^{\,(1)}+\alpha n^{(0)}_\br$.  Here, $\alpha$ is strategically chosen from ($0,1$) to `throttle' the iterative process in order to help maintain convergence and avoid runaway solutions.  These steps are repeated with the new trial density distributions and iterated until the local density converges to within $10^{-6}$ average variation in the density at each site between successive iterations. 

\subsection{Numerical evaluation of the Gross-Pitaevski equation}
The time-evolution of the initial Hamiltonian's equilibrium state after the quench is obtained by numerically integrating the respective time-dependent GP equation. Specifically, Eqn.~\ref{Time} is discretized into small time steps $Jdt$ that were typically about $\sim10^{-5}$.  The time evolution of the initial state is then determined using a fourth-order Runge-Kutta method. In order to ensure convergence, this process is repeated many times with increasingly fine discretization to confirm that there are negligible differences between the solutions. As another check, the total energy and particle number are computed to confirm that they remain constant throughout the time evolution. Eventually, every $1000$ steps the density profile and other observables are computed using the wave function at that instant of time.          

\subsection{Simplifying the GP equation for quenching of the checkerboard flux superfluid}

Setting the GP wavefunction to be $A(t) + \eta_\br B(t)$,
we can substitute this into the full time dependent GP equation to obtain equations of motion for the complex
coefficients $A(t) and B(t)$. To obtain the equation for the staggered density $\Delta n_{AB}$ after the quench, it proves
simpler to define the following variables:
\bea
\Delta n_{AB} (t)  &=& 2 [A^*(t) B(t) + A(t) B^*(t)] \\
{\cal K}(t) &=& |A(t)|^2 - |B(t)|^2 \\
{\cal J}(t) &=& i [A^*(t) B(t) - A(t) B^*(t)].
\eea
Here ${\cal K} is proportional to the bond kinetic energy and ${\cal J} is proportional to the bond current. We then obtain the equations
\bea
\frac{d \Delta n_{AB}}{d t} &=& 4 \tilde{\gamma_0} {\cal K} + 4 \tilde{\epsilon_0} {\cal J}, \\
\frac{d {\cal K}}{d t} &=& - \tilde{\gamma_0} \Delta n_{AB} - U {\cal J} \Delta n_{AB}, \\
\frac{d {\cal J}}{d t} &=& - \tilde{\epsilon_0} \Delta n_{AB} + U \Delta n_{AB} {\cal K},
\eea
where we have suppressed the time label for clarity.
Going to second order in time for $\Delta n_{AB}$ yields
\be
\frac{d^2 \Delta n_{AB}}{d t} = - 4 {\tilde{\lambda}}^2_0 \Delta n_{AB} - 4 U ({\cal J} \tilde{\gamma_0}
- {\cal K} \tilde{\epsilon_0}) \Delta n_{AB},
\ee
where ${\tilde{\lambda}}^2_0 =  \tilde{\gamma_0}^2 + \tilde{\epsilon_0}^2$.
To make progress, we resort to the following approximation, which is valid at early times where we expect
well defined oscillations of $\Delta n_{AB}(t)$.  We replace ${\cal K}$ and ${\cal J}$ by their initial values obtained from
$A(0)$ and $B(0)$ that correspond to their equilibrium, pre-quench, values. Let us call these ${\cal K}_0$ and
${\cal J}_0$. Then we find
\bea
\!\! \frac{d^2 \Delta n_{AB}}{d t} \!\! &\approx&\!\! - \left[ 4 {\tilde{\lambda}}^2_0 + 4 U ({\cal J}_0 \tilde{\gamma_0}
- {\cal K}_0 \tilde{\epsilon_0}) \right] \Delta n_{AB}.
\eea
This yields $\frac{d^2 \Delta n_{AB}}{d t} \approx - \Omega^2 \Delta n_{AB}$, where, 
using the explicit values of ${\cal K}_0$
and ${\cal J}_0$, we obtain
\bea
\Omega^2 = 4 (\tilde\lambda_0^2 + \frac{U n_0}{|\lambda_0|} (\tilde{\epsilon}_0 \epsilon_0 + \tilde{\gamma}_0 \gamma_0))
\eea
Since we start at time $t=0$ with a uniform superfluid having no density modulations, the solution to this takes
the form $\Delta n_{AB}(t) = r \sin \Omega t$. To find $r$, we use the initial rate of change,
\be
(d\Delta n_{AB}/dt)_{t=0} = r \Omega = 4 (\tilde{\gamma_0} {\cal K}_0 + \tilde{\epsilon_0} {\cal J}_0).
\ee
This can be simplified to $r \Omega = 4 {\cal I} (\delta J /J)$, where ${\cal I}$ is the magnitude of the initial 
equilibrium current, which is the same on all bonds, and $\delta J$ is the amount by which we quench the
$x$-bond hopping, leading to the final result
\be
\Delta n_{AB}(t) \approx 4 \frac{\cal I}{\Omega} \frac{\delta J}{J} \sin(\Omega t)
\ee
For a weak quench, $\delta J \ll J$, while a strong quench entails setting $\delta J = - J$. The strength of the quench determines not only the amplitude of the oscillations, but also their frequency $\Omega$.
We find that this result fits very well the early oscillations of $\Delta n_{AB}$ obtained by a direct numerical solution of
the GP equations not only in the continuum but also in the central region of the trap for weak as well as strong
quenches. At later times, the dephasing
of the oscillations in the trap leads to a decay of the $\Delta n_{AB}$ arising from spatial variations of $\Omega$
via its density dependence.

\subsection{`Thermal' noise} 
By imprinting random phase fluctuations on the initial state, we increase the energy of the system. We expect that this
noise will lead to an
excess energy density that will scale as $\Delta E\sim \delta\theta^2_{\rm max}$.  To check this scaling, we computed the
ratio of the excess energy for two different values of $\delta\theta_{\rm max}$, choosing $\delta\theta^{(1)}_{\rm max}=0.5$
and $\delta\theta^{(2)}_{\rm max}=1.0$, and find
$\sqrt{\Delta E^{(2)}/\Delta E^{(1)}}=1.9$, which matches closely to the expected value of $\delta\theta^{(2)}_{\rm max}/\delta\theta^{(1)}_{\rm max}=2$. If we time-evolve this initial state (without making a quench), we expect this excess
energy to lead to a typical state from a `thermal ensemble'  --- detailed issues regarding thermalization will be discussed 
elsewhere. To provide a crude estimate of the effective temperature of this `thermal state' before the quench, we
assume that the dominant excitations in the system induced by such random imprinted phase fluctuations
are the low energy linear sound modes. For $U \ll
J$, the low energy Bogoliubov sound mode in the presence of staggered flux \cite{hemmerich.stagsf} may
be approximated as $\hbar \omega_\bk \approx c \bk$, with the sound speed $c \approx \sqrt{n U/m^*}$. Here, $n$ is
the density, and the inverse effective mass is $1/m^* = 2Ja^2\cos(\phi/4)$ (where $a$ is the lattice
constant). Computing the excess energy density in the center of the trap, we can estimate the
temperature of the `thermal state' as $\delta E = [\zeta(3)/\pi c^2]T^3$.  For our parameters ($\sim 20$ atoms per well, $U\!=\!0.2J$ and $\phi\!=\!\pi/2$, and averaged over states with different initial
randomness), we find the temperatures in the two cases to be $T^{(1)} \approx 3.4J$ and $T^{(2)} \approx
5.2J$, which are both significantly smaller than the Berezinskii-Kosterlitz-Thouless transition
temperature, which can be roughly estimated to be $T_{\rm BKT} \approx \pi n/ 2 m^* \approx 40 J$. This
is consistent with our assumption that only low energy sound modes are excited in the thermal state.

\subsection{Quench induced density dynamics for the staggered flux state of fermions}
We begin with the staggered flux Hamiltonian in momentum space,
\bea
H_{\rm sf}  = {\sum_{\bk}}^{\prime} \Omega_\bk
\Psi^\dg_\bk
(\cos\theta_\bk \tau^z + \sin\theta_\bk \tau^y)
\Psi^\pdg_\bk,
\eea
where $\bQ\equiv (\pi,\pi)$, $\Psi^\dg_\bk = (f^\dg_\bk, f^\dg_{\bk+\bQ})$, and $\tau^{y,z}$ are Pauli matrices. The prime on the momentum sum implies that only momenta in the reduced Brillouin zone are included. Here, we have defined
$\Omega_\bk \! = \! \sqrt{\varepsilon^2_\bk + \gamma^2_\bk}$, $\cos\theta_\bk \! = \! 
\varepsilon_\bk/\Omega_\bk$,
and $\sin\theta_\bk \! = \! \gamma_\bk/\Omega_\bk$, with
\bea
\varepsilon_\bk & = & - 2 (J \cos k_x + J_y \cos \frac{\phi}{2} \cos k_y) \\
\gamma_\bk & = &  - 2 J_y \sin \frac{\phi}{2} \cos k_y.
\eea 
This leads to mode energies $\pm \Omega_\bk$ in the initial state.
A translationally invariant quench of the hopping (say $J_x$) 
ensures that different momentum pairs $(\bk,\bk+\bQ)$ stay decoupled from each other. Nevertheless, this quench
instantaneously changes 
$\varepsilon_\bk \to \tilde{\varepsilon}_\bk$, and 
$\gamma_\bk \to \tilde{\gamma}_\bk$, so that we modify
$(\Omega_\bk,\theta_\bk) \to (\tilde\Omega_\bk,\tilde\theta_\bk)$.

Let us define the initial quasiparticle operators $\alpha_{1,2}$ and the final quasiparticle operators $\beta_{1,2}$ via
\bea
\begin{pmatrix}  \sin(\theta_\bk/2) & \cos(\theta_\bk/2) \\
-i \cos(\theta_\bk/2) & i \sin(\theta_\bk/2) \end{pmatrix}
\begin{pmatrix} \alpha^\pdg_{\bk,1} \\ \alpha^\pdg_{\bk,2} \end{pmatrix}
= 
\begin{pmatrix} f^\pdg_{\bk} \\ f^\pdg_{\bk + \bQ} \end{pmatrix}
\eea
and,
\bea
\begin{pmatrix}  \sin(\tilde\theta_\bk/2) & \cos(\tilde\theta_\bk/2) \\
-i \cos(\tilde\theta_\bk/2) & i \sin(\tilde\theta_\bk/2) \end{pmatrix}
\begin{pmatrix} \beta^\pdg_{\bk,1} \\ \beta^\pdg_{\bk,2} \end{pmatrix}
= 
\begin{pmatrix} f^\pdg_{\bk} \\ f^\pdg_{\bk + \bQ} \end{pmatrix}.
\eea
Here, the quasiparticle $\alpha_{\bk,1}$ ($\alpha_{\bk,2}$) of the initial Hamiltonian has energy $-\Omega_\bk$ ($+\Omega_\bk$), while the quasiparticle of the final Hamiltonian $\beta_{\bk,1}$ ($\beta_{\bk,2}$) has energy $- \tilde \Omega_\bk$ ($+\tilde\Omega_\bk$). For simplicity, let us assume that we are at a filling of less than 
one fermion per two sites, so that only some of the $\alpha_1$ quasiparticles are occupied initially, while none of the $\alpha_2$ quasiparticle states are occupied (although this is easily generalizable to greater fillings). 

We can first transform this into the $\beta$-basis to get the dynamics via
\bea
\!\!\!\! f^\dg_\bk \!\! &=& \!\! \sin(\tilde\theta_\bk/2) {\rm e}^{-i \tilde\Omega_\bk t} 
\beta^\dg_{\bk,1} \! +\!  \cos(\tilde\theta_\bk/2) {\rm e}^{i \tilde\Omega_\bk t}  \beta^\dg_{\bk,2} \label{eq:f1}\\
\!\!\!\! f^\dg_{\bk+\bQ} \!\! &=& \!\! i \cos(\tilde\theta_\bk/2) {\rm e}^{-i \tilde\Omega_\bk t} 
\beta^\dg_{\bk,1} \! -\!  i \sin(\tilde\theta_\bk/2) {\rm e}^{i \tilde\Omega_\bk t}  \beta^\dg_{\bk,2} \label{eq:f2}
\eea
To compute the expectation values, we then need to transform back to $\alpha_{1,2}$ quasiparticles, keeping in mind that the ground state at $t=0$
has no $\alpha_2$ quasiparticles. This means that it suffices to set $\beta_{\bk,1} = \alpha_{\bk,1} \cos(\theta_\bk-\tilde\theta_\bk)/2$ and $\beta_{\bk,2} = \alpha_{\bk,1} \sin(\theta_\bk-\tilde\theta_\bk)/2$.

To compute the density modulation between the two sublattices at a subsequent time, $\Delta n_{AB} (t) \equiv  (n_A-n_B)$, we write
\bea
\Delta n_{AB} (t) &=& \frac{2}{M} \sum_\br (-1)^{x+y} \la f^\dg_\br f^\pdg_\br \ra_t \nonumber\\
&=& \frac{2}{M}
{\sum_\bk}^{\prime} (\la f^\dg_\bk f^\pdg_{\bk+\bQ} \ra_t + \la f^\dg_{\bk+\bQ} f^\pdg_\bk  \ra_t)
\eea
where $M$ is the number of lattice sites.
Using Eq.~(\ref{eq:f1}) and Eq.~(\ref{eq:f2}), we find
\be
\Delta n_{AB} (t) = \frac{2}{M} {\sum_{\{\bk\}_{\rm occ}}}^{\!\!\!\prime} \sin(\theta_\bk - \tilde\theta_\bk) \sin(2 \tilde\Omega_\bk t)
\ee
where the momentum sum runs only over {\it initially occupied} states in the reduced Brillouin zone.

\end{document}